# High current well-directed beams of super-ponderomotive electrons for laser driven nuclear physics applications.

O. N. Rosmej[1,2], M. Gyrdymov[2], M. M. Günther[1], N. E. Andreev[3,4], P. Tavana[2], P. Neumayer[1], S. Zähter[1,2], N. Zahn[2], V. S. Popov[3,4], N. G. Borisenko[5], A. Kantsyrev[6], A. Skobliakov[6], V. Panyushkin[6], A. Bogdanov[6], F. Consoli[7], X. F. Shen[8], A. Pukhov[8]

1) GSI Helmholtzzentrum für Schwerionenforschung GmbH, Planckstr.1, 64291 Darmstadt, Germany
2) Goethe University Frankfurt, Max-von-Laue-Str. 1, 60438 Frankfurt am Main, Germany
3) Joint Institute for High Temperatures, RAS, Izhorskaya st.13, Bldg. 2, 125412 Moscow, Russia
4) Moscow Institute of Physics and Technology (State University), Institutskiy Pereulok 9, 141700 Dolgoprudny Moscow Region, Russia
5) P. N. Lebedev Physical Institute, RAS, Leninsky Prospekt 53, 119991 Moscow, Russia
6) Institute for Theoretical and Experimental Physics named by A. I .Alikhanov of NRC "Kurchatov Institute", B. Cheremuschkinskaya 25, 117218 Moscow, Russia
7) ENEA. Fusion and Nuclear Safety Department, C.R. Frascatti, Via Enrico Fermi, 45, 00044 Frascati, Italy
8) Heinrich-Heine-University Düsseldorf, Universitätsstraße 1, Düsseldorf, Germany

E-mail: o.rosmej@gsi.de

We report on new findings in a laser driven enhanced electron beam generation in the multi MeV energy range at moderate relativistic laser intensities and their applications. In our experiment, an intense sub-picosecond laser pulse propagates through a plasma of a near critical electron density (NCD) and direct laser acceleration (DLA) of electrons takes place. The breakthrough toward high current relativistic electron beams became possible due to application of low density polymer foams of sub-mm thickness. In foams, the NCD-plasma was produced by a mechanism of super-sonic ionization. Compared to NCD-plasmas generated by laser irradiation of conventional foils, the DLA acceleration path in foams was strongly enhanced. Measurements resulted into 11÷13 MeV of the effective electron temperature and up to 100 MeV maximum of the electron energy measured in the laser pulse propagation direction. The growth of the electron energy was accompanied by a strong increase of the number of super-ponderomotive electrons and a well-defined directionality of the electron beam that propagates in a divergence cone with a half angle of 12±1°. For the energy range above 7.5 MeV that is relevant for gamma-driven nuclear reactions, we estimate a charge carried by these well-directed electron beams as high as 50 nC and a corresponding efficiency of the laser energy conversion into electrons of 6%. The electron spectra generated by the DLA-mechanism in NCD-plasma at $10^{19}$ Wcm$^{-2}$ laser intensity were compared with those measured in shots onto conventional metallic foils at ultra-relativistic laser intensities of $10^{21}$ Wcm$^{-2}$. In the last case, the twice lower effective electron temperature and the twice lower maximum of the electron energy were registered. The substantial difference in the electron spectra for these two cases presented itself in the isotope production yield. We observed high yield of nuclear reactions demanding high energy MeV photons and neutrons only in the shots onto the pre-ionized polymer foam layers. Good agreement between the experimental data and the results of Particle-in-Cell and GEANT4 simulations was demonstrated.

## I.  INTRODUCTION

Laser-driven relativistic electron beams are excellent tools for the generation of ultrashort MeV gamma [1-4] and neutron sources [5], THz [6-8] and betatron [9-13] radiation. In the case of well-directed high current beams of relativistic electrons one can reach extreme high luminosity of gamma and neutron sources and use them for radiographic applications [14, 15], laser driven nuclear physics [16], and production of radioisotopes for nuclear medicine [17, 18].

Two mechanisms of laser-driven acceleration are currently being discussed as promising for the generation of high energy electrons in near- and sub-critical plasmas. The first one is the laser wake field acceleration (LWFA) [19]: the intense laser pulse drives strong plasma waves that can trap and accelerate electrons. The most prominent case of LWFA is the so-called bubble regime [20-23]. The LWFA allows to reach the highest electron energies of 10 GeV [24] and finds its applications in high energy physics and potentially in XFEL devices [25-27]. The LWFA works best in tenuous, very under-dense plasmas and ultra-short laser pulses, shorter than the plasma wavelength.

The second mechanism is the direct laser acceleration (DLA) in a plasma channel created by a relativistic laser pulse [28]. In this case, the electron acceleration occurs in the presence of strong quasi-static electric and magnetic fields generated in plasma [31]. Ponderomotive expulsion of background plasma electrons from the channel caused by a relativistic laser pulse creates a radial electrostatic field and at the same time, the current of accelerated electrons generates the azimuthal magnetic field [29-31]. A relativistic electron trapped in the channel experiences transverse betatron oscillations and gains energy efficiently from the laser pulse when the frequency of the betatron oscillations becomes resonant with the Doppler shifted laser frequency [29]. Depending on the plasma density and laser intensity, direct laser acceleration (DLA) at the betatron resonance, stochastic heating [32], transition to wake field acceleration or a combination of these mechanisms is realized.

The DLA works efficiently in near critical density (NCD) plasmas and for sub-ps laser pulses like PHELIX at GSI [33]. Different from LWFA, the DLA does not generate electrons at very high energies, rather, it produces ample amounts of electrons with Boltzmann-like distributions carrying mega-ampere currents. The effective temperature of these distributions can reach several tens of MeV. This huge current of hot electrons is perfectly suited for applications in nuclear physics. Interaction of these relativistic electrons with high Z materials causes MeV gamma-radiation that can drive nuclear reactions resulting in neutron production [5]. This scheme based on the laser accelerated electron beam is one of the important pillars of the laser driven nuclear physics program at ELI-NP [16].

Up to now, only a few experiments were performed to demonstrate the advantages of the discussed mechanism of the electron acceleration. The energy-transfer from an ultra-intense laser pulse with intensity of $10^{20}$ Wcm$^{-2}$ to hot electrons in NCD plasmas depending on the pre-plasma scale length was investigated in [34]. In this experiment, a one-dimensional expansion of the plasma with a well-controlled scale length was produced by a separate ns laser pulse. In the experiment, the coupling of the energy of the ultra-intense laser pulse into hot electrons was analyzed indirectly using measurements of the Cu Kα-intensity and proton spectra. The energy distribution of energetic electrons was not measured, but simulated using a 2D Particle-in Cell (PIC) code. A discovered one order of magnitude variation in the coupling efficiency of the laser energy into fast electrons was explained by the existence of a density gradient optimum that ensures strong laser pulse self-focusing and channeling processes. In [35], measurements of electrons accelerated by a relativistic laser pulse propagating across a mm-long extended under-dense plasma plume with (0.02 ÷ 0.05) $n_{cr}$ were reported. The critical electron density is defined as $n_{cr} = m\omega_L^2/(4\pi e^2)$, where *m* and *e* are the mass of electron at rest and its charge and $\omega_L$ is the laser frequency. The experiment showed a strong increase of the effective temperature and a number of supra-ponderomotive electrons caused by the increased length of a relativistic plasma channel. New results on the electron acceleration from by an ASE pre-pulse pre-ionized foam layers conducted at the Omega EP-laser were reported in [31]. Foam layers of 250 μm thickness and 3 up to 100 mg cm$^{-3}$ ($n_e$ = (0.9 ÷ 30)×10$^{21}$ cm$^{-3}$) mean densities were used as targets. The 1 kJ, 8 ÷ 10 ps short laser pulse had (5.3 ± 1.8) × 10$^{19}$ Wcm$^{-2}$ intensity. An approximation of the high energy tail of the measured electron spectra with a Maxwellian-like function resulted in an effective electron temperature, averaged over several shots, of 6÷10 MeV. Shots onto foam layers with densities above 10$n_{cr}$ did not show a visible effect toward an increased effective electron temperature. The drawback of this experiment is that the intensity and the duration of the ASE pre-pulse were not adapted to the variety of the used foam areal densities.

The production of a hydrodynamically stable NCD-plasma layer remains an important issue. A low density polymer aerogel [38] is a very prospective material for the creation of sub-mm long NCD-plasmas and efficient electron acceleration. In [39], polymer foams of 300-500 μm thickness and 2 mg cm$^{-3}$ mean volume density were used as targets. In foams, the NCD-plasma was produced by a mechanism of super-sonic ionization [40, 41] when a well-defined separate ns-pulse was sent onto the foam target forerunning the relativistic main pulse. The required intensity of the ns-pulse (~5x10$^{13}$ Wcm$^{-2}$) was estimated according to reference [40] in order to reach velocity of the super-sonic ionization wave propagating in a 500 μm long polymer foam, equal to 250μm/ns. The created plasma had an electron density of ~7×10$^{20}$ cm$^{-3}$ or 0.7$n_c$. The application of sub-mm thick low-density foam layers provided a substantial increase of the electron acceleration path in a NCD-plasma compared to the case of freely expanding plasmas created in the interaction of the ns-laser pulse with solid foils. In experiments described in [38], a sub-ps laser pulse of the moderate relativistic intensity interacted with standard metallic foils or pre-ionized low-density foam layers. The effective temperature of super-thermal electrons raised from 1.5-2 MeV, in the case of the interaction with a metallic foil at high laser contrast, up to 13 MeV for the laser shots onto long-scale NCD-plasmas. The high conversion efficiency by the laser interaction with sub-mm-long NCD-plasmas was confirmed by results of the PIC-simulations [36, 39]. In [36] it was shown that up to 80% of the laser energy is converted into electrons, while up to 7% was absorbed by the C, O and H-ions.

In the experiments reported here, we demonstrated a high potential of the super-ponderomotive electrons for applications in the field of laser-driven nuclear physics. High current beam of relativistic electrons generated at ~$10^{19}$ Wcm$^{-2}$ laser intensity in the NCD-plasma was used to generate bremsstrahlung radiation by penetration of high Z samples, where a high yield of gamma-driven nuclear reactions was observed. The reaction yields in this case were higher than in the case of electron beams produced in shots onto standard metallic foils at ultra-relativistic laser intensity of $10^{21}$ Wcm$^{-2}$. Our measurements showed a well-defined directionality of the super-ponderomotive electron beam that was reflected by the angular dependence of the gamma-driven nuclear reaction yields. Comparison between the experimental data and the results of the 3D-PIC simulations of the electron beam parameters and the GEANT4 results on the interaction of the relativistic electrons with high Z materials, demonstrated very good agreement. This allows for a further optimization of the experimental set-up toward a record value of neutron flux above $10^{18}$ cm$^{-2}$s$^{-1}$ by changing only the activation sample geometry.

The paper is organized as follows: laser and target parameters together with the used experimental set-up are described in Section 2; Experimental results on characterization of super-ponderomotive electron beams are presented in Section 3; in Section 4, results of PIC-simulations accounting for the experimental geometry are compared to the experiment; in Section 5, application of well-directed ultra-relativistic electron beams for MeV-gamma and neutron production is discussed; Section 6 summarizes the results.

## II. EXPERIMENTAL SET-UP

Experiments were performed at the Petawatt High Energy Laser for Ion eXperiments (PHELIX) at the Helmholtzzentrum GSI Darmstadt [33] at the highest ns laser contrast $\geq 10^{11}$. A s-polarized laser pulse of 1.053 µm fundamental wavelength delivered by the Nd:glass laser was sent onto targets at 5÷7 degree to the target normal. Two different focusing off-axis parabolic mirrors were used providing peak laser intensities of $(1 \div 2.5) \times 10^{19}$ Wcm$^{-2}$ ($a_L = 2.7 \div 4.27$) and $(7 \div 10) \times 10^{20}$ Wcm$^{-2}$ ($a_L = 22.6 \div 27.0$). Here $a_L$ is the normalized vector potential that scales as $a_L^2 = 0.73 I_{L,18} \lambda^2$ with the laser intensity $I_{L,18}$ normalized to $10^{18}$ Wcm$^{-2}$ and the laser wavelength $\lambda$, in µm. The duration of the laser pulse was $750 \pm 250$ fs. In the case of a moderate relativistic laser intensity $(1 \div 2.5) \times 10^{19}$ Wcm$^{-2}$, $90 \pm 10$ J laser energy measured after the main amplifier was focused by means of a 150 cm off-axis parabolic mirror into an elliptical focal spot with FWHM diameters $12 \pm 2$ µm and $18 \pm 2$ µm containing a laser energy of $E_{FWHM} \simeq (17 \div 22)$ J. In the case of the ultra-relativistic laser intensity, 180 J laser energy was focused into a $2.7 \pm 0.2$ µm × $3.2 \pm 0.2$ µm focal spot by a 40 cm off-axis parabolic mirror. The laser energy in the focal spot was 20% of that after the main amplifier and reached $E_{FWHM} \simeq (36 \div 40)$ J. The laser spot size on target and the laser energy in the focal spot were controlled in every shot. Experiments on the direct laser acceleration of electrons in plasmas of near critical density were performed using the mentioned off-axis parabolic mirror with a focal length of 150 cm. In the case of ultra-high laser intensity, shots were done only onto standard metallic foils.

Cellulose triacetate (TAC, $C_{12}H_{16}O_8$) layers of 2 mg cm$^{-3}$ volume density and 300 ÷ 400 μm thickness [42, 43] were used as targets [39]. A sub-mm long NCD-plasma was produced by sending a well-defined ns pulse forerunning the relativistic main pulse onto a foam. The intensity of the ns laser pulse was kept at ~ 5x10$^{13}$ Wcm$^{-2}$ level in order to initiate a super-sonic ionization wave propagating with 2×10$^7$ cm s$^{-1}$ velocity, for more details see [39, 40]. In the case of fully ionized $C_{12}H_{16}O_8$ atoms, the mean ion charge reaches the value of Z = 4.2 that corresponds to an electron density of 0.65×10$^{21}$ cm$^{-3}$. This value is slightly lower than the critical density for $\lambda$ = 1.053 μm ($n_{cr}$ = 10$^{21}$ cm$^{-3}$). The ns-pulse was focused on target by the same parabolic mirror as the short pulse. Generation of a long scale NCD-plasma requires focusing optics with hundreds of micrometers long Rayleigh length that was the case for the 150 cm off-axis parabolic mirror. This ensures rather constant ns laser pulse intensity along the whole layer thickness. The delay between the peak of the ns pulse and the relativistic main pulse was fixed by 2÷3ns. Experimental set-ups and a target holder with a low-density foam layer fixed inside a 2.5 mm in diameter Cu-washer are shown in [FIG.1].

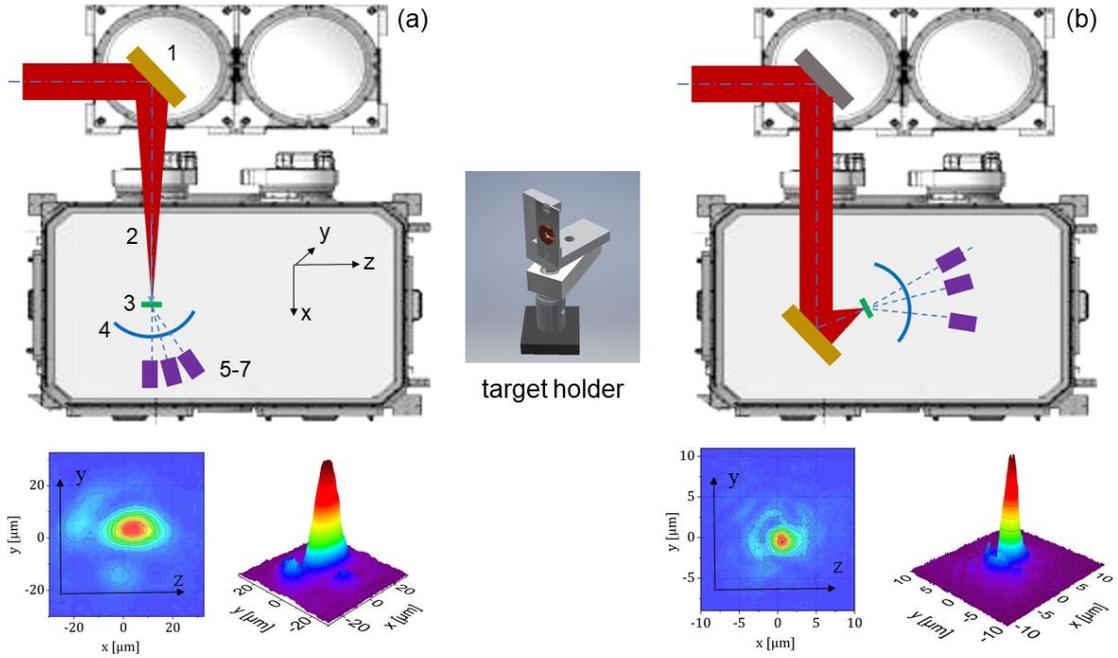

FIG. 1. Experimental set-ups (top-view) used for shots (a) at moderate and (b) ultra-high relativistic intensities. Scheme (a) shows an off-axis focusing parabola (1), laser beam (2), target position (3), stack of cylinders (4), and three magnet spectrometers (5÷7). The laser intensity distributions for both set-ups are also shown.

In both set-ups, the electron spectra were measured simultaneously at three different angles with respect to the laser axis. The spectrometers were equipped with 0.99 T static magnets and imaging plates (IP) for the detection of the electron signal [44, 45]. The spectral resolution of the spectrometers with a 300 μm (width) × 1000 μm (height) entrance slit was numerically simulated using measured 2D B-field distribution. It was shown that an experimental error in the energy measurements caused by the 300 μm entrance slit is not higher than 2%. This allows for reliably measuring of electron spectra from 1.75 up to 100 MeV. The spectrometers were placed in the horizontal plane *XOZ* (which is perpendicular to the laser polarization along *OY*) at a distance of 405 mm from the interaction point around the target at 0°, 15° and 45° [5-7 in FIG. 1(a)]. Massive 20 up to 40 mm long WCu - collimator blocks with 3 mm entrance hole were placed in front of every 0.99 T spectrometer in order to shield the front plate from gamma-rays and to increase the signal-to-noise ratio. The angular distribution of the electron beam in a wide range of angles was measured by means of a stack of three stainless steel cylindrical plates of 3 mm thickness each, rather similar to that described in [46]. The cylinder-stack had a curvature radius of 200 mm and was placed 230 mm away from the target position [4 in FIG.1(a)]. The observation angle was 0° ± 50° in the horizontal direction and 0° ± 15° in the vertical direction, where 0° corresponds to the laser axis. A horizontal 3 mm wide slit centered at the laser pulse height allowed electrons to propagate to the magnet spectrometers placed behind the cylinder-stack. In order to map the position of the electron beam in space, small holes with a 20 mm interval in vertical and horizontal directions were drilled into the front plate. Large area IPs were placed between the first and second, and the second and third cylindrical plates to map the spatial distribution of electrons with $E > 3$ MeV and $E > 7.5$ MeV correspondently.

A nuclear activation based diagnostic [47, 48] was used to characterize MeV-gamma and neutron beams generated by the interaction of the super-ponderomotive electrons with high-Z materials. The activation samples, consisting of stacked together Au, Ta, and Cr plates, were fixed at horizontal angles of 6°±1° and 16°±1° to the laser pulse axis [FIG. 2(b, d)]. After irradiation by electrons and gammas, all activation samples were counted multiple times on a low background HPGe-detector to identify nuclides via known γ-ray energies, intensities, and half-life times. Reaction yields of the obtained isotopes $^{196,194,192}$Au, $^{180,178m}$Ta and $^{51,49}$Cr produced due to photodisintegration were used for reconstruction of the MeV bremsstrahlung spectrum. A 0.25mm thin In-foil was placed between Ta and Cr for measurements of the neutron yield [49].

### III. EXPERIMENTAL RESULTS

Figure 2 shows raw electron spectra measured in two selected laser shots at 0° with respect to the laser axis by means of the 0.99 T electron spectrometer and angular distribution of the electron beam with $E > 3$ MeV and $E > 7.5$ MeV measured using the cylinder-stack. For registration of the electron signal Fuji BAS IP MS type was used as detector, while for diagnostic of the electron angular distribution Fuji BAS IP TR. For each picture in [FIG. 2], the maximum intensity of the related raw signals expressed in PSL (Photostimulated Luminescence) is indicated. The presented shots were made at ~$10^{19}$ Wcm$^{-2}$ laser intensity onto a 10 µm thin Au-foil [FIG. 2(a, b)] and a pre-ionized foam combined with a 10 µm Au-foil attached to the rear side [FIG. 2(c, d)]. In the case of the laser shot onto the pre-ionized foam, the raw electron spectrum shows a 10-fold enhanced IP-signal [compare FIG. 2 (a, c)] that reflects a corresponding increase of the number of electrons with energies above 2 MeV [44]. In the case of the foil target, the maximum of the measured electron energy lays in the area up to 15 MeV, while for shots onto the pre-ionized foam stacked with the Au-foil the maximum electron energy reached the value of 95÷100 MeV. The experimental error in the energy definition is caused mostly by the 300 µm width of the spectrometer entrance slit and is estimated to be not higher than 2%.

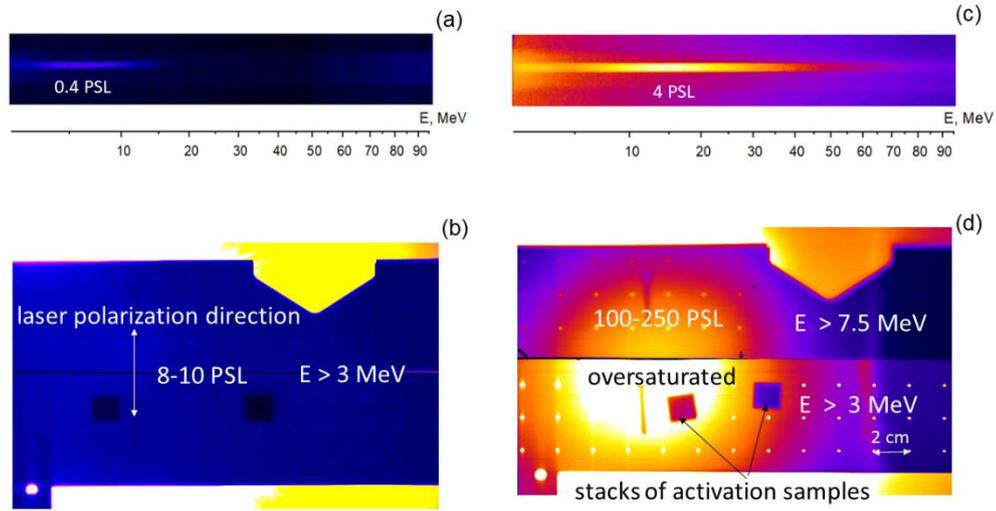

FIG. 2. Raw electron spectra measured along the laser axis by means of the 0.99 T electron spectrometer (a, c) and angular distribution of the electron beam with E > 3 MeV and >7.5 MeV registered using the cylinder-stack (b, d): (a, b) shot at $1.6\times10^{19}$ Wcm$^{-2}$ laser intensity onto a 10 µm thin Au-foil; (c, d) shot at $1.5\times10^{19}$ Wcm$^{-2}$ laser intensity onto pre-ionized foam layer of 325 µm thickness combined with a 10 µm Au-foil.

At the same time, for the shot onto the pre-ionized foam, we observed by means of the cylinder-stack a strong collimation of electrons with energies E > 3 MeV (first IP) and E > 7.5 MeV (second IP) into a well-directed electron beam with a half of a divergence angle of 12°±1° at FWHM [FIG. 2(d)]. The IP signal produced by electrons with energies higher than 3 MeV that could pass through the first 3mm of stainless steel was oversaturated (> 300 PSL), while the IP signal obtained for the laser shot onto foil reached only 8 ÷ 10 PSL [FIG. 2(b)].

GEANT4 [49-51] simulations were performed to define the input of electrons and X-rays into the IP-image. The discussed experimental geometry [FIG. 1(a)], the measured electron energy distribution [FIG.4] and the BAS IP TR imaging plate response to electrons and photons [44] were used as input parameters. Simulations showed that the contribution of photons to the IP signal is less than 5% and that the obtained signal can be attributed mainly to the electron angular distribution.

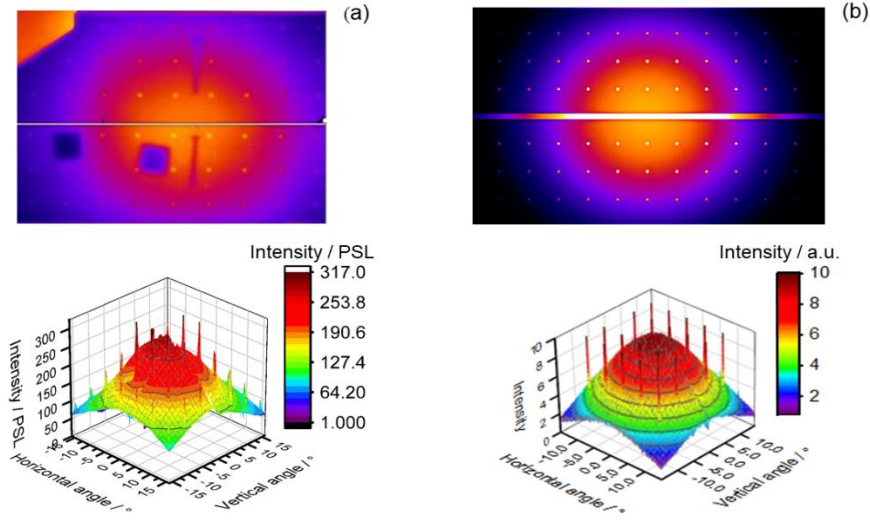

FIG. 3. Comparison of the (a) IP image obtained in experiments using the cylinder-stack and (b) GEANT4 simulation result. "Needles" in the 3D-representation of the image are caused by the holes drilled in the first steel cylinder at a distance of 2 cm from each other to map the electron distribution spatially.

The electron energy distribution measured by three spectrometers placed in a horizontal plane *XOZ* at 0°, 15° and 45° to the laser pulse propagation direction [FIG. 1(a)] are shown in [FIG. 4]. The shot was made at $1.5 \times 10^{19}$ Wcm$^{-2}$ laser intensity onto a 325 µm thick foam layer pre-ionized by the ns pulse.

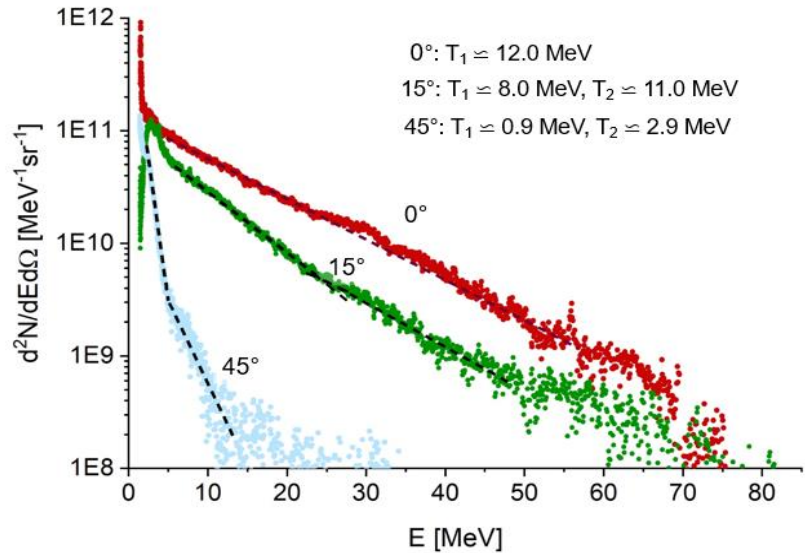

FIG. 4. Energy spectra of electrons per steradian measured at 0°(red), 15°(green) and 45°(blue) to the laser propagation direction for a shot onto the pre-ionized foam layer at $1.9 \times 10^{19}$ Wcm$^{-2}$ laser intensity.

One can see that the majority of electrons is accelerated in the laser axis direction (0°) as it is expected from the DLA-process. The measured electron spectra were approximated by a Maxwellian-like distribution functions with one or two temperatures. The electron energy distribution measured in the laser pulse propagation direction (0°) is approximated by one effective temperature $T_1 = 12 \pm 1.4$ MeV. The temperature and the number of accelerated electrons drops down to $T_1 \simeq 8.0$ MeV ($T_2 \simeq 11.0$ MeV) at 15° and further to $T_1 \simeq 0.9$ MeV ($T_2 \simeq 2.9$ MeV) at 45°.

In the experiment, besides foam targets, we used a combination of foams with 10 µm thin foils or 1-2mm thick metallic plates attached to the foam layers from the rear side. Spectra measured at 0°, 15° and 45° in the interaction of the $1.9 \times 10^{19}$ Wcm$^{-2}$ laser pulse with the foam layer stacked together with the 10 µm thin Au-foil [FIG. 2(b) and FIG. 5] show a strong angular dependence similar to that shown in [FIG. 4]. Additionally, one observes high energy (E > 25 MeV) electron tails at 0° and 15° that can be described by an exponential function with a very high second effective temperature of $T_2 = 28 \pm 3.4$ MeV for 0° and $T_2 = 19 \pm 2.3$ MeV for 15°. This part of the spectrum contains a very high fraction of electrons and plays an important role in generation of tens of MeV gamma-radiation. The effect of metallic foils attached to the rear side of the foam layers is a subject of our further experimental and numerical investigations. The spectra of ponderomotive electrons with effective temperatures between $1.4 \pm 0.12$ MeV (45°) and $2.2 \pm 0.15$ MeV (0°) obtained in the interaction of $1.6 \times 10^{19}$ Wcm$^{-2}$ laser pulse with the 10 µm Ti-foil show an angular isotropy. [FIG. 5].

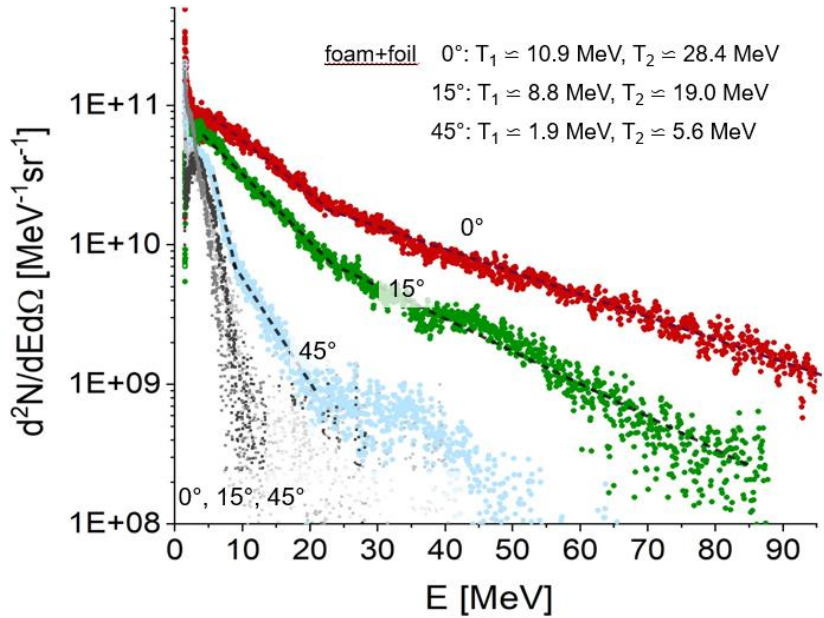

FIG. 5. Electron spectra measured at 0°(red), 15°(green) and 45°(blue) to the laser pulse axis for shots at ~$10^{19}$ Wcm$^{-2}$ laser intensity. Shots were made onto the pre-ionized foam layer combined with 10 µm Au-foil and directly onto the 10 µm Ti-foil (gray).

Measurements of the electron energy distribution were also performed inside the electron divergence cone by means of three 0.99 T spectrometers placed at angles of 0° and ± 7° to the laser axis [FIG. 6].

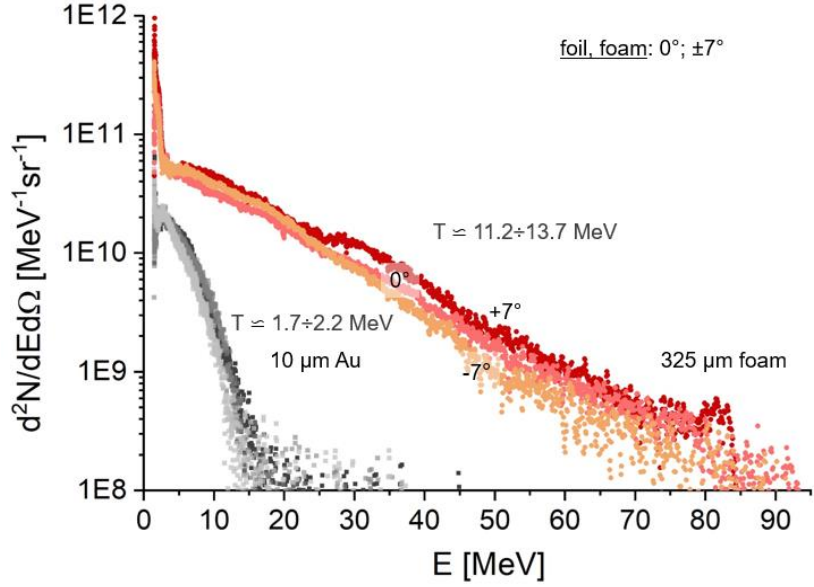

FIG. 6. Electron energy distribution measured inside the divergence cone of the super-ponderomotive electrons in laser shots onto a 10 µm thin Au-foil (gray) at $1.6\times10^{19}$ Wcm$^{-2}$ laser intensity and a pre-ionized low-density foam layer of 325 µm thickness (red) at $2.5\times10^{19}$ Wcm$^{-2}$ laser intensity. Spectra were measured by means of three spectrometers at 0° and ±7° to the laser axis.

No noticeable difference in the effective electron temperatures was detected for all three directions neither in the case of the foam target nor in the case of the Au-foil. For shots onto pre-ionized foam layers the effective temperatures lay between 11.2±1.3 MeV and 13.7±1.6 MeV, while for laser shots onto the Au-foil between 1.7±0.2 and 2.2±0.26 MeV.

In this experimental campaign, the electron energy distribution was also measured in the interaction of ultra-relativistic laser pulses of $10^{21}$ Wcm$^{-2}$ intensity [FIG. 1(b)] with metallic foils.

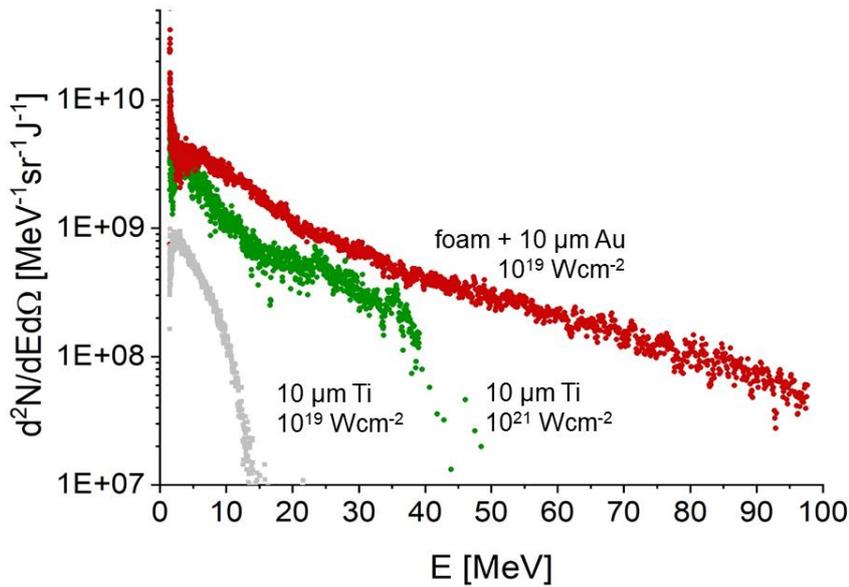

FIG. 7. Electron energy distributions is normalized to the laser energy inside FWHM of the focal spot in Joules. Shots were made onto the pre-ionized low density foam layer in combination with 10 µm Au-foil at $1.5\times10^{19}$ Wcm$^{-2}$ laser intensity (red) and onto 10µm thin Ti-foils irradiated by the $9\times10^{20}$ Wcm$^{-2}$ (green) and $1.6\times10^{19}$ Wcm$^{-2}$ (gray) laser intensities. In all cases, spectra were measured along the laser axis (0°).

Figure 7 shows electron spectra measured at 0° in shots made onto the pre-ionized low density foam layer in combination with 10 µm Au-foil at $1.5\times10^{19}$ Wcm$^{-2}$ laser intensity and onto 10µm thin Ti-foils irradiated at $9\times10^{20}$ Wcm$^{-2}$ and $1.6\times10^{19}$ Wcm$^{-2}$ laser intensities. The number of accelerated electrons presented in [FIG.7] is normalized to the laser energy contained in the FWHM of the focal spot, since in the case of the short focusing parabolic mirror this energy was twice higher than in the long-focus case (36÷40 J vs. 17÷20 J). The effective temperature of the major fraction of electrons in the case of the ultra-relativistic laser intensity is estimated as high as $6.74 \pm 1.2$ MeV, which is lower than for shots onto pre-ionized foams at $10^{19}$ Wcm$^{-2}$ (~ 11÷13 MeV). Another important difference one can observe analyzing the maximum of the measured electron energy. In contrast to shots with the ultra-relativistic laser intensity, where the maximum of the measured electron energy does not exceed 40 MeV, the electron spectra, measured at 0° by irradiation of foams in combination with thin foils, show a very stable signal up to 90÷100 MeV electron energy [FIG. 5, 7].

Once the optimal parameters for the ns-pulse that drives the super-sonic ionization inside the foam were found, the DLA mechanism in the relativistic plasma channel reaches a high level of reproducibility. In measurements at 0°, the effective temperature of the main fraction of super-ponderomotive electrons, averaged over 10 shots onto the pre-ionize foam layers only and onto the combination of foams and foils, reaches $11.5\pm2.0$ MeV. Figure 8 summarizes data on the number of electrons in different energy ranges and the maximum of the measured electron energy for selected shots made onto various targets irradiated at two laser intensities: $1.5\div1.9\times10^{19}$ Wcm$^{-2}$ (blue) and $9.0\times10^{20}$ Wcm$^{-2}$ (red-dashed) and measured at 0°, 15° and 45°. Data is normalized to laser energy in the selected shot. The experimental error of ~ 25% that occurs by the evaluation of the electron number is caused mostly by the uncertainty of the IP response to electron impact. This uncertainty remains constant for all electron energies above 0.1 MeV [44]. Additionally, the IP signal fading [45] was taken into account. In the experiment, the IPs were scanned firstly 30-50 min after each laser shot, so that for the signal corrections a fading factor of 0.65÷0.7 was used.

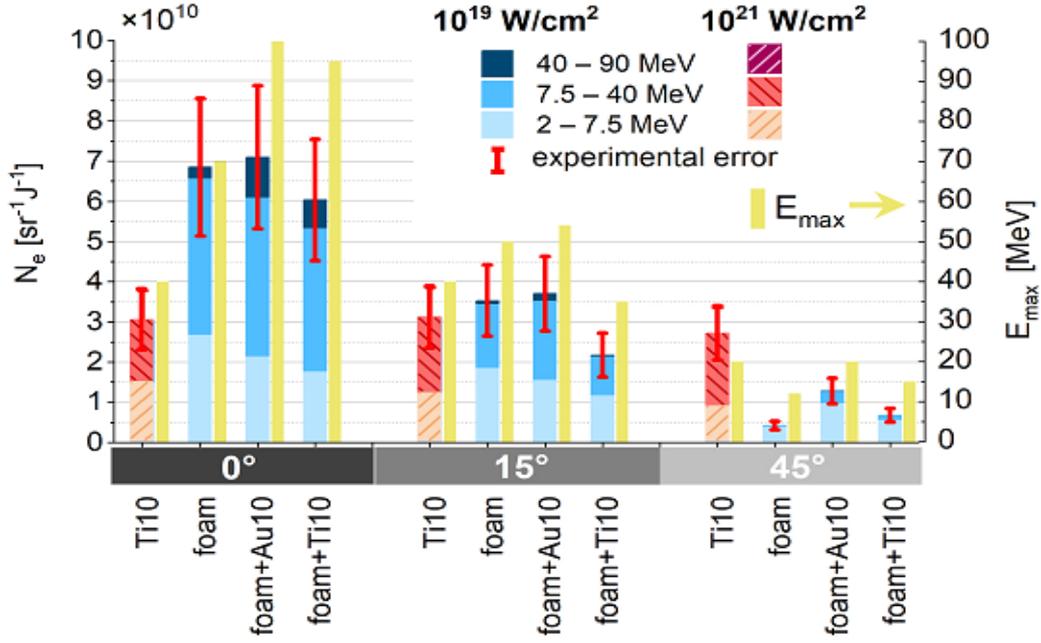

FIG. 8. Number of electrons $N_e$ per steradian and per J laser energy for different electron energy ranges for selected shots onto various target types irradiated at two laser intensities: 1.5÷1.9×10¹⁹ Wcm⁻² (blue) and 9×10²⁰ Wcm⁻² (red, dashed) and measured at 0°, 15° and 45°. The maximum of the measured electron energy $E_{max}$ is also shown.

The full columns height in [FIG. 8] corresponds to the total amount of electrons with energies E > 2 MeV in steradian normalized to the laser energy contained in the FWHM of the focal spot. For laser intensities of 1÷2.5 10¹⁹ Wcm⁻² this energy was of 17 ÷ 22 J, while for ultra-relativistic intensity of 10²¹ Wcm⁻² it achieves 37 ÷ 40 J. The discussed uncertainty in the electron number caused by the IP response is valid in every energy range presented in [FIG. 8]. In the diagram, this uncertainty is shown for the total number of electrons. In the case of the foam target, one counts 7×10¹⁰ sr⁻¹ J⁻¹ electrons with E > 2 MeV and 4.4×10¹⁰ sr⁻¹ J⁻¹ electrons with E > 7.5 MeV. In the case of the foam layer stacked together with the Au-foil, the total number of electrons with E > 2 MeV is close to the foam-case, while the number of electrons with E > 7.5 MeV is slightly higher and reaches the value of 5.0×10¹⁰ sr⁻¹ J⁻¹. The darkest colored column´s part represents electrons with E > 40 MeV with the highest number obtained in the case of the combination of the foam layer with the 10μm thin Au-foil. This part of the electron spectrum is of interest e.g. in respect to (γ, 5n) reactions in Au with the gamma ray threshold at 38.7 MeV. In results obtained at relativistic laser intensity of 9×10²⁰ Wcm⁻² in the shot onto Ti-foil this part is missing [FIG. 8]. Although the role of the thin foils in the combination with foams demands further theoretical and experimental investigations, their positive effect is clearly seen by comparing the values of the measured maximum electron energy. In the cases of "foam + Ti" and "foam + Au" [FIG. 8] they lay between 95 and 100 MeV, instead of 70 MeV for the case of only foam.

The measured number of electrons in the electron energy range 2÷10 MeV has to be corrected at least by a factor 2 due to loss of particles caused by the collimator and the tight entrance slit of the spectrometer. In some measurements, a 0.25 T spectrometer was placed at 18° to the laser axis additionally to the 0.99 T one that was placed at 15°. This was done in order to understand the influence of the 20mm long WCu collimators used in combination with the 0.99 T spectrometers on the measured electron number. Another goal was to register electrons with energies below 1.75 MeV, which were not accessible for the 0.99 T field. The imaging plate of the 0.25 T spectrometer registered in total twice more electrons with energies between 2 and 10 MeV compared to the 0.99T spectrometer. Probably this factor is even higher for spectra detected at 0°. At electron energies higher than 10 MeV this difference vanished. We suppose that the reason for this effect is a quasi-static electric field created at the collimator front plate by the high current relativistic electrons that reached the collimator at the very beginning of the interaction [50]. Electrons that approach the collimator at later time can be slightly deviated from the initial trajectory by this field. After propagation over 20 mm inside the collimator and an additional 20 mm gap until the spectrometer front, some fraction of electrons can miss the 300 µm tight entrance hole. This effect was very weak in shots onto thin foils, probably due to ~ 10-times lower number of accelerated electrons.

## IV. PIC-SIMULATIONS

3D PIC simulations were performed using the Virtual Laser Plasma Laboratory (VLPL) code [51] for the laser parameters and interaction geometry used in the experiment. In particular, a laser pulse intensity in time and space was approximated by the Gaussian distribution. Elliptical form of the focal spot was taken from the experiment with FWHM axes 11µm in a vertical and 15 µm a horizontal direction. The laser pulse energy in the FWHM focal spot of 17.5 J and the FWHM pulse length of 700 fs resulted into the laser intensity of $2.5 \times 10^{19}$ Wcm$^{-2}$ with $a_L = 4.28$. The homogeneous plasma was composed of electrons and fully ionized ions of carbon, hydrogen and oxygen. Simulations accounted for the ion type and the ion fraction in accordance with the chemical composition of triacetate cellulose $C_{12}H_{16}O_8$, see e.g. [42, 43]. The simulation box had sizes of $350 \times 75 \times 75$ µm$^3$. The first 10 and the last 15 µm from the total 350 µm of the space in $x$-direction (the laser axis) were free of the plasma initially. Sizes of a numerical cell were 0.1 µm along the $x$-axis and 0.5 µm along the $y$-axis and the $z$-axis. The number of particles per cell in the simulation were 4 for the electrons and 1 for the ions of each type. Boundary conditions were absorbing for particles and fields in each direction.

The initial electron density profile (together with the neutralizing ion density) at the moment of the main pulse arrival was a step-like with the electron density $n_e = 0.65\ n_{cr}$. Previously, PIC-simulations were performed for step-like density profile with $n_{cr}$ and 0.5 $n_{cr}$ [36, 37] and for a partially ramped density profile in order to account for plasma expansion toward the main laser pulse [39]. The simulations result in a very similar overall behavior of the energy and angular distributions of super-ponderomotive electrons in all mentioned cases.

Figure 9 gives the 3D view of the laser intensity by propagation of the laser pulse in the NCD plasma at time moments $t_1 = 100$ fs (a) and $t_2 = 433$ fs (b). The laser pulse propagates from the left and at $t = 0$ its maximum intensity is on the target front side ($x = 10$ µm). We observe strong self-focusing (a) and filamentation (b) of the laser pulse that produces the main channel and a few side channels inside the plasma.

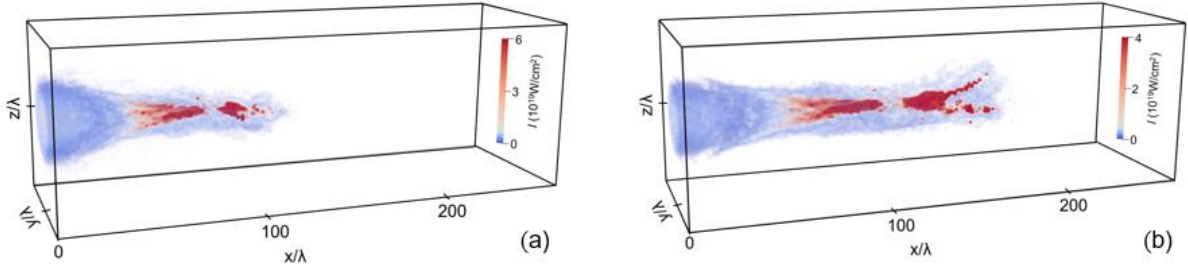

FIG. 9. 3D view of the laser intensity distribution at time moments $t_1 = 100$ fs (a) and $t_2 = 433$ fs (b). The time $t = 0$ corresponds to the maximum laser pulse intensity on the target front side (x = 10 μm).

The laser ponderomotive force expels background plasma electrons off the channel and creates a radial quasi-electrostatic field. At the same time, the current of the accelerated relativistic electrons generates an azimuthal magnetic field. The structure of the quasi-static electric and magnetic fields was investigated analytically in [28-30] and numerically for conditions relevant to our experiment in [36, 39]. Electrons are trapped inside the plasma channel and experience betatron oscillations in the combined quasi-static channel fields. They can get in resonance with the laser field and be strongly accelerated [28, 29].

Figure 10(a) shows a longitudinal momentum $p_x$ at $t = 100$ fs that grows along the laser pulse propagation direction due to the DLA-process. The high values of the transversal electron momentum $p_y \sim$ 30-50 $mc$ (polarization direction) [FIG. 10(b)] and $p_z \sim$ 20 $mc$ [FIG. 10(c)] are much larger than the normalized amplitude of the laser pulse $a_0 = eE_0/mc\omega \sim 5$. These are the electron betatron oscillations in the plasma channel quasi-static fields. The large value of the electron transverse momenta is due to the resonance between the electron betatron frequency and the Doppler shifted laser frequency [28, 29]. Figure 10(d) zooms the $p_y$ vs $x$ electron phase space in the range 40÷70 $x/\lambda$. We clearly see that the electron transverse oscillations are modulated at the laser phase. The huge amplitude of the oscillations $p_y \sim$ 30÷50 $mc$ is due to the resonance between the laser pulse and the electron betatron frequency in the channel fields.

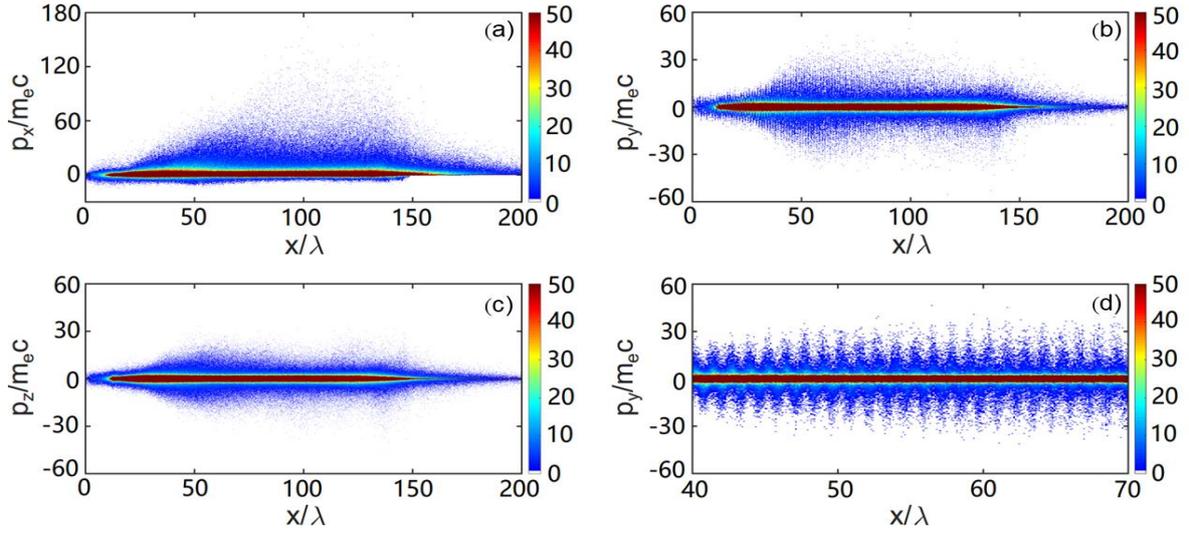

FIG. 10. Snapshots of the electron phase space 100 fs after the laser pulse peak intensity arrived at the left plasma boundary: (a) momentum $p_x$ vs $x$; (b) momentum $p_y$ vs $x$; (c) momentum $p_z$ vs $x$; (d) zoomed part of momentum $p_y$ vs $x$ in the range $x/\lambda = [40, 70]$.

In the PIC simulation, we register electrons leaving the NCD plasma. Angular distribution of the electrons with $E > 7.5$ MeV is presented in [FIG. 11] in spherical coordinates with a polar axis $OX$ along the laser propagation direction: $\theta = \arctan(\sqrt{p_y^2 + p_z^2}/p_x)$, $\varphi = \arctan(p_z/p_y)$.

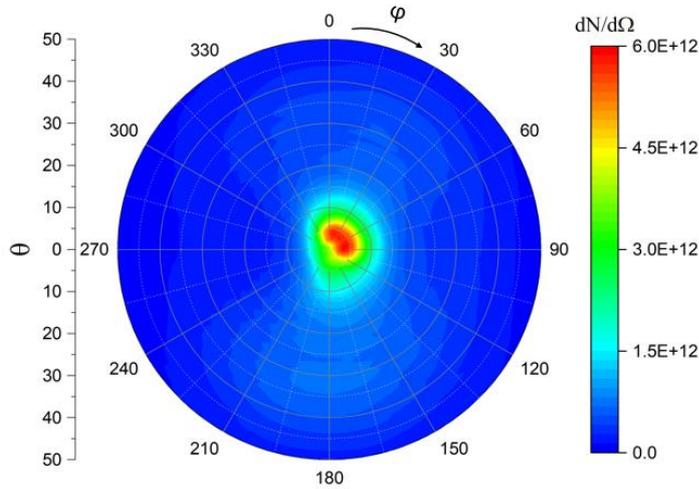

FIG. 11. Angular distributions of electrons $dN/d\Omega$ [sr$^{-1}$] with energies E > 7.5 MeV that left the simulation box by the time $t = 2.5$ ps

One can see that a high fraction of the super-ponderomotive electrons is accelerated in the laser direction and propagate in the rather tight divergence cone with a half angle of $10°\div12°$. This result supports the experimental measurements presented in [FIG. 3].

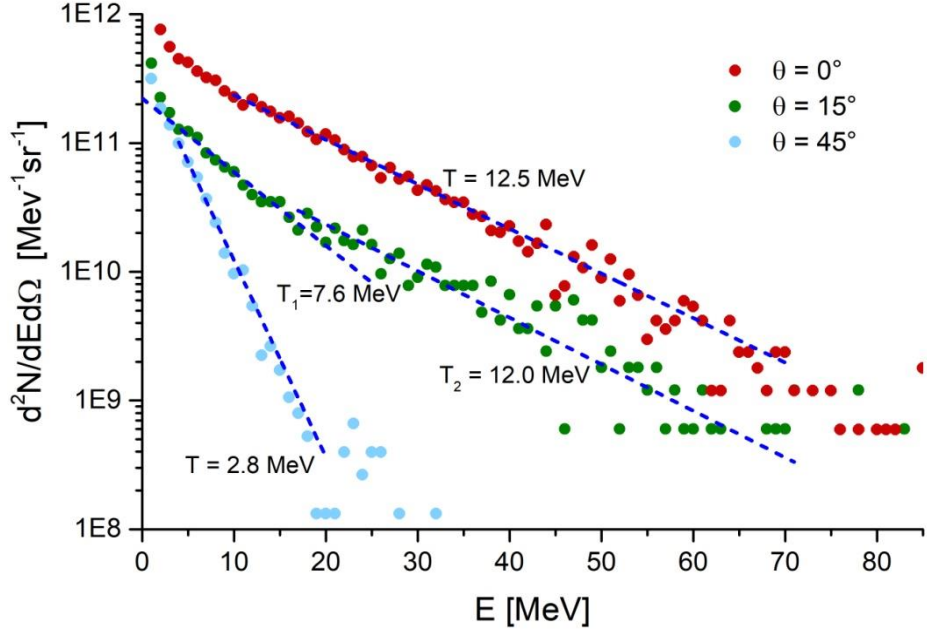

FIG. 12. Energy distributions of electrons per steradian ($d^2N/dEd\Omega$) that leave the simulation box at $t = 2.5$ ps for three different angles $\theta = 0°$(red), 15°(green) and 45°(blue) in the horizontal plane *XOZ*.

The 3D capability of the PIC code allows for simulations that are close to real experimental conditions. Thus, the absolute energy spectra, i.e., the number of accelerated electrons in any energy range and their angular distribution can be obtained. The electron spectra were simulated in the horizontal plane for $\theta = 0°\pm2°$ (red dots), 15°$\pm2°$(green dots) and 45°$\pm2°$ (blue dots), which correspond to the positions of the electron spectrometers in the experiment [FIG.1(a)]. The obtained effective temperatures are in good agreement with the experimental results presented in [FIG. 4] for all three observation directions. According to simulations, the total number of electrons with E > 2 MeV propagating in $2\pi$ reaches the value of $5\times10^{12}$ providing 27% conversion of the laser energy into electrons. Current simulations show, that only 12% of electrons with E > 2 MeV are contained in the 0.16 sr divergence cone. Another situation is with the super-ponderomotive electrons with energies beyond 7.5 MeV that are relevant for gamma-driven nuclear reactions. Simulations result into $N_e = 3.3\times10^{11}$ electrons contained in 0.16 sr, that is 30% of all super-ponderomotive electrons with E > 7.5 MeV propagating in $2\pi$. This number is in good agreement with the experimental value of $2\pm0.6\times10^{11}$ that can be slightly corrected upwards due to discussed electron loss at the collimator entrance. Based on these numbers, one can estimate a charge carried by the super-ponderomotive electrons propagating with E > 7.5 MeV in 0.16 sr divergence cone as high as ~ 50 nC. For these electrons, the corresponding conversion efficiency reaches ~ 6% or ~ 3 nC per J laser energy contained in the focal spot.

## V. DISCUSSION

Highly charged and well directed electron beams with energies of several MeVs are excellently suited for generation of ultra-intense gamma and neutron sources by penetration of high Z materials [5, 47, 48]. In [5] authors reported on a novel compact laser-driven neutron source with an unprecedented short pulse duration (<50 ps) and record peak flux (>$10^{18}$ n/cm$^2$/s). In their experiment, high-energy electron jets were generated from a thin (<3 μm) plastic foil firstly heated with a nanosecond pulse and then irradiated, with a 60 ns delay, by a relativistic laser pulse of 90 J energy on target and peak vacuum intensity of 5÷7×$10^{20}$ Wcm$^{-2}$. These intense electron beams propagated in a 1.8 cm thick Cu-converter and generated MeV bremsstrahlung and neutrons in gamma-driven nuclear reactions. This scheme based on laser accelerated electron beams is one of the important pillars of the laser driven nuclear physics program at ELI-NP [16]. In [18], it was numerically demonstrated that the interaction of the super-ponderomotive electron beam with a mm-thick high Z foil leads to efficient production of $^{62-64}$Cu isotopes that are important for medical applications. All of these open a great prospective for application of the laser driven relativistic electrons for generation of an ultra-high neutron flux >$10^{20}$ neutrons/cm$^2$/s that is required to drive r-process nuclear synthesis in laboratory conditions.

In our experiment, 1 mm-thick activation samples (Au, Ta, Cr) together with a 0.25 mm thin Indium plate were placed at 6°±1° to the laser pulse propagation direction 150÷230 mm away from the target [FIG. 2(c, d)]. Neutrons were generated predominantly via (γ, xn) reactions in all these materials. In shots at 1÷2×$10^{19}$ Wcm$^{-2}$ laser intensity onto low-density foam layers stacked together with mm thick high-Z converter targets, we detected at 5°-7° to laser pulse propagation direction, up to 40-times higher $^{178}$Ta and $^{194}$Au isotopes yields with threshold gamma energies E$_\gamma$ ≃ 22÷23 MeV, compared to the laser shots directly onto the convertor plate at $10^{21}$ Wcm$^{-2}$ laser intensity. The $^{192}$Au isotopes created in (γ, 5n) reaction with the cross section threshold at E$_\gamma$ ≃ 38.7 MeV and Indium isotopes $^{112m}$In, $^{112}$In and $^{111m}$In that requires high flux of 2.5 MeV and 14 MeV neutrons and 14÷20 MeV gammas were observed only in shots at ~$10^{19}$ Wcm$^{-2}$ onto pre-ionized foams stacked together with 1mm thick Au-plates. The modeling of the interaction processes by means of GEANT4 Monte Carlo code [52-54] was performed for this combined target. In simulations, the experimental geometry and the measured electron energy distribution together with the divergence angle of super-ponderomotive electrons were used as input parameters. The activation samples of Ta, Au, In and Cr were "placed" at 5° to the laser pulse axis 180 mm away from the laser-foam interaction point. For simulations, cross sections for (γ, xn) and (n, xn´) reactions in Au, Ta and Cr were taken from [52]. A cross check between these cross sections and those from [55] used for the reconstruction of the gamma-spectrum from experimentally measured isotope yields showed a satisfactory agreement. The values of the measured and simulated reaction yield of Ta, Au und Cr isotopes per steradian and per 1 Joule laser energy are presented in [FIG. 13]. The normalization was made by taking the solid angle of 7×$10^{-3}$ sr covered by the activation samples into account. The uncertainty in the reaction yields is mainly due to the uncertainty in the number of counts in the full energy peaks. For determination of the reaction yield, the efficiency of the detector, the intensities of γ-lines and the dead time of detector have been taken into account. Blue columns in [FIG.13] present the reaction yields obtained in the shot at 1.5×$10^{19}$ Wcm$^{-2}$ laser intensity onto a pre-ionized foam layer stacked together with 1 mm thick Au-plate used as a radiator. Green columns show reaction yields simulated for this case with the GEANT 4 code. Good agreement between experimentally measured and simulated values is clearly seen. The reaction yields measured in the shot at $10^{21}$ Wcm$^{-2}$ directly onto the radiator plate are presented by sand colored columns.

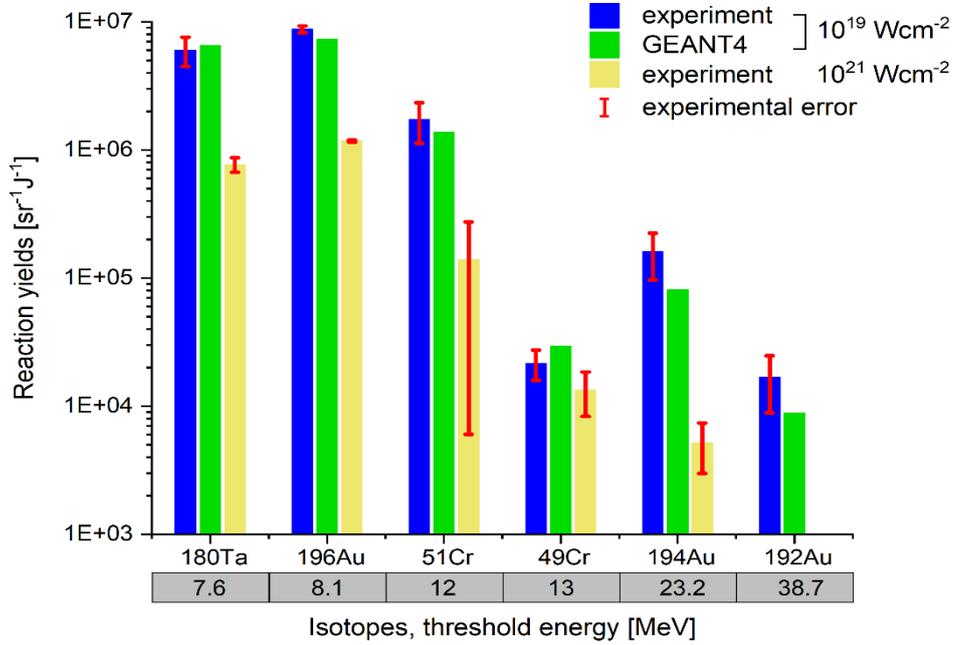

FIG.13. Experimental data on (γ, xn) reaction yields in Au, Ta und Cr and results of GEANT4 simulations normalized to steradian and the laser energy in the corresponding shot. Blue columns present reaction yields obtained in the shot at 1.5×10$^{19}$ Wcm$^{-2}$ laser intensity onto a pre-ionized foam layer stacked together with 1 mm thick Au-plate used as a radiator. Green columns show reaction yields simulated for this case with GEANT 4. The reaction yields measured in the shot at $10^{21}$Wcm$^{-2}$ directly onto the radiator plate is shown by sand colored columns. In the gray row below the figure, the threshold photon energies for the corresponding atomic element are presented.

The simulations showed that the penetration of the super-ponderomotive electrons through 1mm thick Au, Ta and Cr-samples causes bremsstrahlung radiation with temperatures of $T_1 \simeq 7.3$ MeV ($E_\gamma = 10 \div 25$ MeV) and $T_2 \simeq 11.2$ MeV ($E_\gamma = 25 \div 60$ MeV). The number of photons with E ≥ 7.7 MeV that can initiate disintegration in Au, Ta and Cr was up to 1.6×10$^9$ cm$^{-2}$ (5.1×10$^{11}$ sr$^{-1}$). At these photon energies, the gamma radiation is directed along the beam of super-ponderomotive electrons. Simulation results are in good agreement with the gamma-spectrum reconstructed from the measured yields of Au, Ta and Cr isotopes with an effective temperature T=12.7±2.4 MeV and a gamma fluence $N_\gamma$ (>7 MeV) = 2.0±0.5×10$^9$ cm$^{-2}$ or ~ 4.5×10$^{11}$ sr$^{-1}$.

The neutron spectrum evaluated from the experimentally measured In-isotope production was approximated by an exponential function with an effective temperature of 1.2±0.6 MeV and a neutron fluence of 4.1±2.1×10$^7$ cm$^{-2}$ (~10$^{10}$ sr$^{-1}$). The GEANT4 simulations resulted in a neutron spectrum with an effective temperature of 1.6 MeV and a total neutron fluence of ~ 10$^7$ cm$^{-2}$ summarized over 15mm × 15mm front and back, and four 3 mm × 15 mm side areas of the sample stack. This difference between the experiment and the simulations can be attributed to the fact that the current version of the GEANT4 does not account for the production of isotopes metastable levels. We would like to emphasize that the above mentioned high gamma and neutron fluences were generated in the samples situated 180 mm away from the laser-foam interaction point. This means, that by taking the solid angle covered by activation samples of $7\times10^{-3}$ sr into account, only 4% of the super-ponderomotive electrons participated in the gamma-ray production.

Good agreement in the isotope reaction yields between the GEANT4 simulations and the experiment allows for the further optimization of the experimental set-up toward record values of the gamma and neutron production. Indeed, the number of MeV photons and neutrons can be strongly enhanced by the optimization of the set-up geometry and the activation material thicknesses in accordance with the measured electron energy distribution. More details on the nuclear activation diagnostics and results of GEANT4 simulations will be published elsewhere by M. M. Günther et al.

## VI. SUMMARY

The novel experimental results and numerical simulations demonstrate an extremely high capability of the well-directed high current relativistic electron beams to be used in novel laser assisted applications exploiting already existing sub PW laser systems of moderate relativistic intensities. Ultra relativistic electron beams were produced in the interaction of $\sim 10^{19}$ Wcm$^{-2}$ laser pulses with plasmas of near critical electron density by the mechanism of direct laser acceleration (DLA). Sub-mm thick low-density polymer foams, pre-ionized by a well-defined ns pulse, provided a long acceleration path for electrons confined in the relativistic plasma channel. The angular dependence of the electron acceleration process was investigated. The highest effective electron temperature of 11÷14 MeV and the maximum of the electron energy up to 70÷100 MeV were measured inside the electron beam divergence cone with the half angle of 12°±1° (0.16 sr). These results are supported by the full 3D PIC-simulations.

The measured effective electron temperature and the maximum of the electron energy were twice higher for shots onto pre-ionized foams at $\sim 10^{19}$ Wcm$^{-2}$ than for direct laser shots onto standard foils at ultra-relativistic laser intensity of $\sim 10^{21}$ Wcm$^{-2}$. The substantial difference in the electron spectra for these two cases became visible in the isotope production yield. We observed high yields nuclear reactions demanding high energy MeV photons and neutrons in shots onto pre-ionized foam layers. The results of GEANT4 simulations of the electron beam interaction with the stack of activation samples, performed for the selected shot, showed a good agreement with the measured gamma-driven nuclear reaction yields. This allows for the further optimization of the experimental set-up toward record values of the photon and neutron fluxes by changing only the activation sample geometry.

## ACNOWLEDGMENTS

The experimental group is very thankful for the support provided by the PHELIX-laser team at GSI-Darmstadt. We thank also Dr. B. Borm for the development of the electron spectrometers. The work was funded by the grant No. 16 APPA (GSI) of the Ministry of Science and Higher Education of the Russian Federation, by the Project DFG PU 213/9-1, the RFBR project 19-02-00875, the RFBR and ROSATOM project 20-21-00150, and by the Euratom research and training program 2019–2020 under the grant agreement No. 633053.